\documentclass[twocolumn,pra]{revtex4}
\usepackage[latin9]{inputenc}
\usepackage{mathrsfs}
\usepackage{amsmath}
\usepackage{amssymb}
\usepackage{graphicx}

\makeatletter
\@ifundefined{textcolor}{}
{%
 \definecolor{BLACK}{gray}{0}
 \definecolor{WHITE}{gray}{1}
 \definecolor{RED}{rgb}{1,0,0}
 \definecolor{GREEN}{rgb}{0,1,0}
 \definecolor{BLUE}{rgb}{0,0,1}
 \definecolor{CYAN}{cmyk}{1,0,0,0}
 \definecolor{MAGENTA}{cmyk}{0,1,0,0}
 \definecolor{YELLOW}{cmyk}{0,0,1,0}
}

\usepackage{amsfonts}
\makeatother

\begin{document}

\title{Incoherent control of electromagnetically induced transparency and
Aulter-Townes splitting}

\author{Chang-Ling Zou, $^{1,2,3}$ Yan-Lei Zhang, $^{1,3}$ Liang Jiang,
$^{2}$ Xu-Bo Zou, $^{1,3,*}$ Guang-Can Guo $^{1,3}$}

\affiliation{$^{1}$Key Laboratory of Quantum Information, University of Science
and Technology of China, CAS, Hefei, Anhui 230026, China}

\affiliation{$^{2}$Department of Applied Physics, Yale University, New Haven,
CT 06511, USA}

\affiliation{$^{3}$Synergetic Innovation Center of Quantum Information \& Quantum
Physics, University of Science and Technology of China, Hefei, Anhui
230026, China}

\email{xbz@ustc.edu.cn}

\begin{abstract}
The absorption and dispersion of probe light is studied in an unified
framework of three-level system, with coherent laser driving and incoherent
pumping and relaxation. The electromagnetically induced transparency
(EIT) and Autler-Townes splitting (ATS) are studied in details. In
the phase diagram of the unified three-level system, there are distinct
parameter regimes corresponding to different lineshapes and mechanisms,
and the incoherent transition could control the cross-over between
EIT and ATS. The incoherent control of the three-level system enables
the investigation of various phenomena in quantum optics, and is beneficial
for experiments of light-matter interactions.
\end{abstract}


\maketitle
\emph{Introduction.-} Great progresses have been achieved in the coherent
light-atom interaction, which is essential for the basic research
and application of quantum physics \cite{Scully1997}. Novel phenomena,
such electromagnetically induced transparency (EIT), have been predicted
and observed in experiments \cite{Boller1991,Harris1997}. In the
medium composed by three-level systems, the dispersion and absorption
of light propagating can be controlled by another laser beam, thus
the EIT have been utilized for slow light and information storage
\cite{Lukin2001,Lukin2003,RevModPhys.77.633}. Recently, the EIT phenomena
have also been generalized to other systems, such as quantum well
\cite{Phillips2003}, optical microcavities \cite{Xu2006}, surface
plasmon \cite{Liu2009}, meta-materials \cite{Papasimakis2008} and
optomechanics \cite{Weis2010}. The sharp transparent window in the
transmission or reflection spectrum is also beneficial for applications
including sensors \cite{Liu2010,Lukyanchuk2010} and switches \cite{Albert2011}.
However, there is a similar phenomenon known as Autler-Townes splitting
(ATS), which shows two distinguished resonances in spectrum. ATS and
EIT are usually confusing since both of them show transparency windows
and split resonances. Thus, efforts have been dedicated to distinguish
ATS and EIT recently \cite{PhysRevLett.107.163604,PhysRevA.81.053836,Zhu2013,Giner2013,Oishi2013,Peng2014}.

In this paper, we proposed a unified framework of the three-level
system (TLS). In the unified framework, the populations of different
energy level can be controlled continuously by incoherent pumping
and relaxation, and the interconversion between different energy level
structures is possible. The mechanisms of EIT and ATS are studied
with incoherent control. It's found that there are distinct regimes
corresponding to different spectrum shapes can be visited by controlling
driving laser power and the incoherent transition rates.

\begin{figure}
\begin{centering}
\includegraphics[width=7cm]{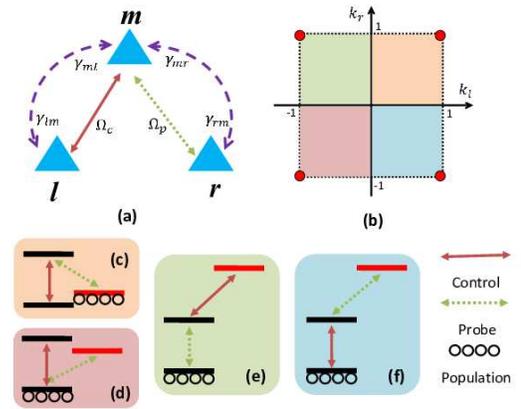}
\par\end{centering}
\protect\caption{(Color online) (a) Schematic illustration of general three-level system,
the levels are denoted by $\left|l\right\rangle $, $\left|r\right\rangle $
and $\left|m\right\rangle $. The coherent and incoherent transitions
are denoted by arrows, but the transition between $l$ and $r$ is
forbidden. (b) The incoherent transition ratio parameter space of
TLS. The four typical categories of energy level structure: $\Lambda$-TLS
(c), $V$-TLS (d), upper-level-driving $\Xi$-TLS (e) and lower-level-driving
$\Xi$-TLS (f), which are corresponding to the four vertexes (red
dots) in (b). }
\end{figure}

\emph{Model.-} Shown in Fig. 1(a) is the schematic illustration of
generalized TLS, with energy levels denoted by left ($\left|l\right\rangle $),
right ($\left|r\right\rangle $) and middle ($\left|m\right\rangle $).
The transition between $l$ and $r$ is forbidden, and there are control
and probe lasers excite the transitions $l-m$ and $m-r$ near-resonantly.
Due to the symmetry, we denote the $l-m$ transition as control laser
and the $m-r$ transition as probe without loss of generality. The
kernel of the unified framework of TLS is that the incoherent transitions
$l-m$, $m-l$, $r-m$ and $m-r$ are all taken into consideration.
These incoherent transitions mainly due to the spontaneous emission
relaxation process and incoherent pumping. It is well known that the
atomic spontaneous emission rate can be inhibited and accelerated
by changing the electromagnetic density of states due to thew Purcell
effect \cite{Scully1997}, thus the relaxation rate can be controlled
in experiment. The incoherent pumping can be realized by incoherent
light source or pumping electrons to energy levels outside the TLS,
and incoherent transition rate can be adjusted simply by changing
the pumping power.

By controlling the incoherent transition rate, the steady state populations
of TLS without external coherent driving at thermal equilibrium can
be arbitrary mixing of the three levels. Denote the incoherent transition
rate from level $a$ to level $b$ as $\gamma_{ba}$, and introduce
the transition ratios
\begin{eqnarray}
k_{l} & = & \frac{\gamma_{lm}-\gamma_{ml}}{\gamma_{lm}+\gamma_{ml}},\\
k_{r} & = & \frac{\gamma_{rm}-\gamma_{mr}}{\gamma_{rm}+\gamma_{mr}}.
\end{eqnarray}
Thus, we can have steady state populations equivalent to four typical
energy level structures {[}as shown in Fig. 1(c)-(f){]}: (i) lower-level-driving
$\Xi$-TLS with $k_{l}=1$ and $k_{r}=-1$; (ii) upper-level-driving
$\Xi$-TLS with $k_{l}=-1$ and $k_{r}=1$, (iii) $V$-TLS with $k_{l}=-1$
and $k_{r}=-1$, (iv) $\Lambda$-TLS with $k_{l}=1$ and $k_{r}=1$.
Previous studies on the coherent light-TLS interactions are focused
on these four specific atomic energy level structures, and found that
EIT can only be observed in $\Lambda$- {[}Fig. 1(c){]} and upper-level-driving
$\Xi$- {[}Fig. 1(e){]} TLSs \cite{RevModPhys.77.633}. In the parameter
space $k_{l},\ k_{r}\in[-1,\ 1]$, the well-studied four typical energy
level structures are corresponding to the vertexes in Fig. 1(b). Interesting
physical phenomena can be observed in the intermediate parameter region
by tuning the incoherent control parameters. For example, increase
$\gamma_{lm}$ monotonously, the effective energy level structures
are transformed continuously from $\Lambda$-TLS to $\Xi$-TLS. Therefore,
the EIT or ATS spectrum can be observed in arbitrary TLS by incoherent
control.

\emph{Hamiltonian and solutions.- }The Hamiltonian describing the
generalized TLS {[}Fig. 1(a){]} is $H=-\Delta_{c}\left|l\right\rangle \left\langle l\right|-\Delta_{p}\left|r\right\rangle \left\langle r\right|+\frac{1}{2}(\Omega_{c}\left|m\right\rangle \left\langle l\right|+\Omega_{p}\left|m\right\rangle \left\langle r\right|+h.c.)$
with control and probe laser detuning $\Delta_{c}$ and $\Delta_{p}$,
coupling strength $\Omega_{c}$ and $\Omega_{p}$. The dynamics of
the system is governed by the Master equation \cite{Scully1997},
which reads
\begin{align}
\frac{d}{dt}\rho= & -i[H,\rho]+\sum_{x=l,m,r}\gamma_{xx}\mathscr{L}(\left|x\right\rangle \left\langle x\right|)\rho\nonumber \\
 & +\sum_{x=l,r}\left[\gamma_{mx}\mathscr{L}(\left|m\right\rangle \left\langle x\right|)\rho+\gamma_{xm}\mathscr{L}(\left|x\right\rangle \left\langle m\right|)\rho\right],\label{eq:master}
\end{align}
with the Lindblad super-operator $\mathscr{L}(o)\rho=o\rho o^{\dagger}-\frac{1}{2}\rho o^{\dagger}o-\frac{1}{2}o^{\dagger}o\rho$.
The parameter $\gamma_{xx}$ denotes the dephasing of $\left|x\right\rangle $
($x=l,m,r$). The steady state solution of the Master equation to
zeroth order perturbation of $\Omega_{p}$ is $\rho_{mm}=\frac{-\gamma_{mr}\left(A-\gamma_{ml}\right)}{-A\left(2\gamma_{mr}+\gamma_{rm}\right)+\gamma_{ml}\left(\gamma_{mr}+\gamma_{rm}\right)+\gamma_{mr}\gamma_{lm}}$,
where $A=\frac{\left|\Omega_{c}\right|^{2}}{2}\mathrm{Re}[\frac{1}{i\Delta_{c}-(\gamma_{ml}+\gamma_{lm}+\gamma_{rm}+\gamma_{ll}+\gamma_{mm})/2}]$.
To the first order of $\Omega_{p}$, the density matrix element $\rho_{mr}=\frac{\Omega_{p}}{2}\chi$,
where the susceptibility of the probe laser can be written as
\begin{equation}
\chi=\frac{iC\left(i\Delta_{p}+D\right)}{\omega_{mr}\omega_{lr}+\frac{\left|\Omega_{c}\right|^{2}}{4}}.\label{eq:susceptibility}
\end{equation}
Here, $C=\frac{\gamma_{rm}-\gamma_{mr}}{\gamma_{mr}}\rho_{mm}$, $D=\frac{\gamma_{ml}+\gamma_{mr}+\gamma_{ll}+\gamma_{rr}}{2}-\frac{\left|\Omega_{c}\right|^{2}}{4\omega_{lm}}\frac{\gamma_{mr}-\left(2\gamma_{mr}+\gamma_{rm}\right)\rho_{mm}}{\left(\gamma_{rm}-\gamma_{mr}\right)\rho_{mm}}$,
$\omega_{lr}=-i(\Delta_{p}-\Delta_{c})-\frac{\gamma_{ml}+\gamma_{mr}+\gamma_{ll}+\gamma_{rr}}{2}$
and $\omega_{mr}=-i\Delta_{p}-\frac{\gamma_{lm}+\gamma_{mr}+\gamma_{rm}+\gamma_{mm}+\gamma_{rr}}{2}$.
Obviously, $C$ and $D$ are real numbers, and the denominator in
Eq. \ref{eq:susceptibility} can be written in the form as $\omega_{mr}\omega_{lr}+\frac{\left|\Omega_{c}\right|^{2}}{4}=\left(i\Delta_{p}+\delta_{1}\right)\left(i\Delta_{p}+\delta_{2}\right)$
with $\delta_{1,2}=\gamma_{s}\pm\frac{\sqrt{\Omega_{s}^{2}-|\Omega_{c}|^{2}}}{2}$.
Here $\gamma_{s}=\frac{\gamma_{ml}+2\gamma_{mr}+\gamma_{lm}+\gamma_{rm}+\gamma_{mm}+\gamma_{ll}+2\gamma_{rr}}{4}$,
and
\begin{equation}
\Omega_{s}=|\frac{\gamma_{lm}+\gamma_{rm}+\gamma_{mm}-\gamma_{ml}-\gamma_{ll}}{2}|
\end{equation}
is the critical control strength.

When $|\Omega_{c}|=\Omega_{s}$, which means that $\delta_{1}=\delta_{2}$,
we have
\begin{equation}
\chi=\frac{iC\left(i\Delta_{p}+D\right)}{\left(i\Delta_{p}+\delta_{1}\right)^{2}}.\label{eq:critical}
\end{equation}

When $\delta_{1}\neq\delta_{2}$, the susceptibility is composed by
two resonances
\begin{equation}
\chi=\frac{iC}{\delta_{2}-\delta_{1}}(\frac{D-\delta_{1}}{i\Delta_{p}+\delta_{1}}-\frac{D-\delta_{2}}{i\Delta_{p}+\delta_{2}}).
\end{equation}
From the view of resonance interference, the TLS in control laser
field can be divided to two regimes: (a) degenerate regime ($\left|\Omega_{c}\right|<\Omega_{s}$),
(b) non-degenerate regime ($\left|\Omega_{c}\right|>\Omega_{s}$).
For the sake of the simplicity and clarity of the results, we set
the control laser detuning $\Delta_{c}=0$. In the following, we are
focus on the incoherent control by adjusting $\gamma_{mr}$. Similar
phenomena can be expected by adjusting other parameters, since the
essential underlying physics is the same.

\begin{figure}
\includegraphics[width=9cm]{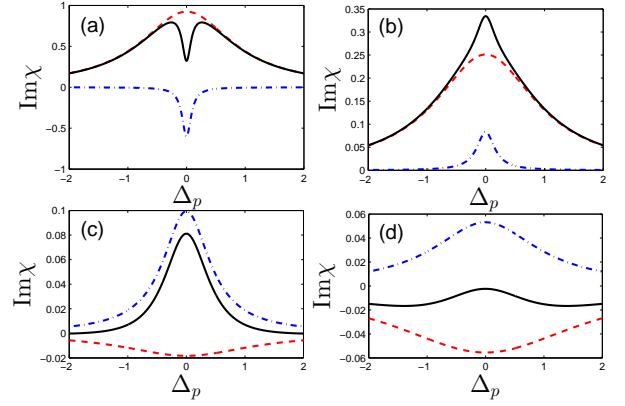}\protect\caption{In the degenerate regime, $\mathrm{Im}\chi$ as a function of the
probe laser detuning $\Delta_{p}$ with the increasing of $\gamma_{mr}$.
Each subplot includes the first resonance $\chi_{1}$ (the red dashed
line), second resonance $\chi_{2}$ (the blue dash-dotted line), and
the total $\mathrm{Im}\chi$ (the black solid line). (a)-(d) The incoherent
transition rates are $\gamma_{mr}=0.02$, $0.2$, $0.8$ and $2$,
with fixed $\Omega_{c}=0.5$. Other parameters are $\gamma_{ml}=0.01$,
$\gamma_{lm}=1$, $\gamma_{rm}=1$, $\gamma_{ll}=0.01$, $\gamma_{mm}=0.01$,
and $\gamma_{rr}=0.01$.}
\end{figure}

\emph{Degenerate Regime.-} When the control light is weak $\left(\left|\Omega_{c}\right|<\Omega_{s}\right)$,
the two resonances are degenerated but with different linewidth $\delta_{1}$
and $\delta_{2}$ $\left(\delta_{1}>\delta_{2}\right)$. Then we can
obtain the imaginary part of linear susceptibility (proportional to
the absorption coefficient)
\begin{equation}
\mathrm{Im}(\chi)=\chi_{1}+\chi_{2},
\end{equation}
 as an overlap of two resonances ($k=1,2$) with
\begin{equation}
\chi_{k}=\frac{C(D-\delta_{k})}{(-1)^{k}(\delta_{1}-\delta_{2})}\mathrm{Re}\left(\frac{1}{i\Delta_{p}+\delta_{k}}\right).
\end{equation}
In the degenerate regime, it is possible to suppress the absorption
through the cancellation of two resonances. We plot $\mathrm{Im}(\chi)$
as a function of the probe laser detuning $\Delta_{p}$ with the increasing
of the incoherent transition rate $\gamma_{mr}$ in the Fig. 2. Each
subplot includes the first resonance $\chi_{1}$ (the red dashed line),
second resonance $\chi_{2}$ (the blue dash-dotted line), and $\mathrm{Im}\chi$
(the black solid line), which clearly show the relation between the
two resonances and the total absorption. From the parameters, we obtain
the critical control strength $\Omega_{s}=0.995$, and the $\Omega_{c}<\Omega_{s}$
is satisfied. In the Fig. 2(a), we set $\gamma_{mr}=0.02$, and the
EIT phenomenon is observed with a very sharp dip in the total absorption
spectrum. We have $C>0,$ $D<\delta_{2}$ and $\delta_{2}\ll\delta_{1},$
and it means that the narrow dip from the second resonance is added
to the first resonance with the wide absorption peak, which is observed
from the resonance $\chi_{1}$ and $\chi_{2}$. Fig. 2(b) shows that
the absorption is enhanced with $\gamma_{mr}=0.2$, and now we have
$C>0$ and $\delta_{2}<D<\delta_{1}$. Obviously the amplitude of
the two resonances are with the same sign, which leads to a sharp
peak of total absorption and is known as electromagnetic induced absorption
(EIA). With increasing $\gamma_{mr}$, we have $D>\delta_{1}$, and
the amplitude becomes negative from the $\chi_{1}$ in the Fig. 2(c).
The strength of the first resonance is weaker compared to the second
resonance, and the total absorption spectrum is only weakened. When
we set $\gamma_{mr}=2>\gamma_{rm}$, which means $C<0$, then we have
$D<\delta_{2}$. We observe the EIT in the amplification spectrum
from the Fig. 2(d), which means that the amplification of probe light
is suppressed by the control laser. From above results, we can only
expect the EIT or EIA in the degenerate regime. It's worth noting
that $\Omega_{s}=0$ when $\gamma_{lm}+\gamma_{rm}+\gamma_{mm}-\gamma_{ml}-\gamma_{ll}=0$,
which means there is no degenerate regime.

\begin{figure}
\includegraphics[width=9cm]{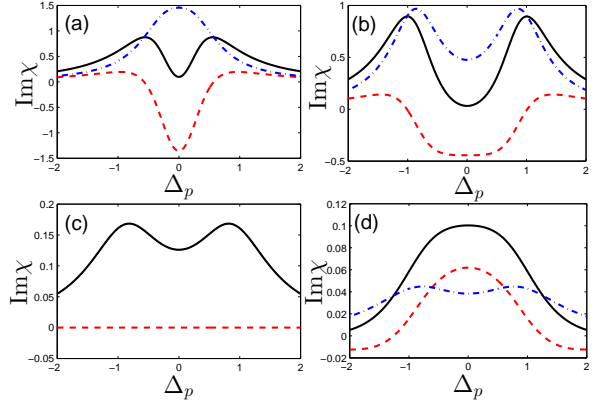}\protect\caption{In the non-degenerate regime, $\mathrm{Im}\chi$ as a function of
the probe laser detuning $\Delta_{p}$ with the different $\gamma_{mr}$
and $\Omega_{c}$. Each subplot includes $\mathrm{Im}(\chi_{E})$
(the red dashed line), $\mathrm{Re}(\chi_{A})$ (the blue dash-dotted
line), and the total $\mathrm{Im}\chi$ (the black solid line). (a)
$\Omega_{c}=1.1$, $\gamma_{mr}=0.02$ . (b) $\Omega_{c}=2$, $\gamma_{mr}=0.02$
.(c) $\Omega_{c}=2$, $\gamma_{mr}=0.5025$ .(d) $\Omega_{c}=2$,
$\gamma_{mr}=0.8$ . The remaining parameters are the same as Fig.
2.}
\end{figure}

\emph{Non-degenerate Regime.-} When pump laser is strong that $\left|\Omega_{c}\right|>\Omega_{s}$,
the frequency splitting between two resonances is given by
\begin{equation}
\triangle_{s}=\sqrt{\left|\Omega_{c}\right|^{2}-\Omega_{s}^{2}},
\end{equation}
and the linewidth of two resonances are equal to $\gamma_{s}$. The
expression of susceptibility is significantly different from the degenerate
regime, where the coefficients of two resonances are both real numbers.
The imaginary part of linear susceptibility for splitting regime can
be written as
\begin{equation}
\mathrm{Im}(\chi)=\mathrm{Im}(\chi_{E})+\mathrm{Re}(\chi_{A}),
\end{equation}
where $\chi_{E}=-\frac{C(D-\gamma_{s})}{\triangle_{s}}\left(\frac{1}{\omega_{+}}-\frac{1}{\omega_{-}}\right),\ \chi_{A}=\frac{C}{2}\left(\frac{1}{\omega_{+}}+\frac{1}{\omega_{-}}\right).$and
$\omega_{\pm}=i\left(\Delta_{p}\pm\frac{1}{2}\triangle_{s}\right)+\gamma_{s}$.

Fig. 3 is a set of $\mathrm{Im}(\chi)$ as a function of the probe
laser detuning $\Delta_{p}$ with the different incoherent transition
rate $\gamma_{mr}$ and the control light $\Omega_{c}$ in the condition
$\left|\Omega_{c}\right|>\Omega_{s}$. The red dashed line represents
$\mathrm{Im}(\chi_{E})$, which is related to the EIT as interaction,
the blue dash-dotted line $\mathrm{Re}(\chi_{A})$ is the sum of two
Lorentzian peaks, corresponding to the ATS, and the black solid line
is the total spectrum $\mathrm{Im}(\chi)$. In Fig. 3(a), we choose
the parameters with $D\ll\gamma_{s}$, $\Delta_{s}<\gamma_{s}$. In
this case, the splitting of two Lorentzian peaks is smaller than the
linewidth, and the dip is only from the $\mathrm{Im}(\chi_{E})$.
The condition $\Delta_{s}>\gamma_{s}$ is observed in the Fig. 3(b)
with increasing of $\Omega_{c}$, and we can see two peaks from the
blue dash-dotted line, which reflects the cross-over from EIT to ATS.
If we tune the incoherent transition rate $\gamma_{mr}$ to satisfy
$D=\gamma_{s}$, $\chi_{E}$ is zero as shown in Fig. 3(c), and the
spectrum becomes ATS consisting of two Lorentzian shaped peaks. When
$D>\gamma_{s}$, as plotted in Fig. 3(d), the transparency from the
splitting of two Lorentz peaks is hided because of the $\mathrm{Im}(\chi_{E})$.
From the Fig. 3, we can realize the cross-over from EIT to ATS by
adjusting incoherent transition rate $\gamma_{mr}$ and the control
light $\Omega_{c}$.

\begin{figure}
\begin{centering}
\includegraphics[width=7cm]{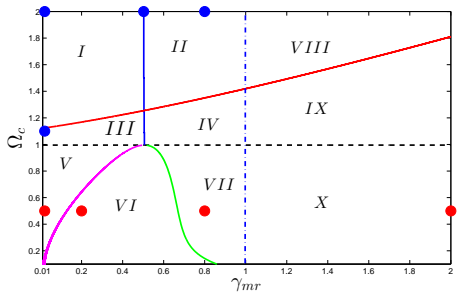}
\par\end{centering}

\protect\caption{The black dashed line : $|\Omega_{c}|=\Omega_{s}$, the blue dash-dotted
line : $C=0$ $\left(\gamma_{mr}=\gamma_{rm}=1\right)$, the blue
line : $D=\gamma_{s},$ the red line: $\Delta_{s}=\gamma_{s},$ the
magenta line: $D=\delta_{2},$ the green line: $D=\delta_{1},$ and
the red and blue dots correspond to Fig. 2 and 3, respectively. The
remaining parameters are the same as Fig. 2.}
\end{figure}

\emph{Phase Diagram.-} To intuitively understand different lineshapes
and the crossover between EIT and ATS, we plot the phase diagram in
the Fig. 4 by controlling incoherent transition rate $\gamma_{mr}$
and the control light $\Omega_{c}$. The black dashed line is critical
condition for splitting $|\Omega_{c}|=\Omega_{s}$, which shows spectrum
from the Eq. \ref{eq:critical}. The blue dash-dotted line shows $C=0$
$\left(\gamma_{mr}=\gamma_{rm}=1\right)$, and for the regions with
incoherent transition rate $\gamma_{mr}<\gamma_{rm}\ (\gamma_{mr}>\gamma_{rm})$,
we have $C>0\ (C<0)$. The red and blue dots denotes parameters of
Fig. 2 and 3, respectivly. Below the black dashed line is the degenerate
regime with $|\Omega_{c}|<\Omega_{s}$, and the magenta and green
lines indicate $D=\delta_{2}$ and $D=\delta_{1}$, respectively.
In the zone $V$, the two resonances are out of phase, which leads
to the EIT shown in the Fig. 2(a). In contrast, the two resonances
are in phase in the zone $VI$, and the spectrum shows EIA, as shown
in the Fig. 2(b). The zone $VII$ and $\uppercase\expandafter{\romannumeral10}$
corresponds to $C>0,$ $D>\delta_{1}$ and $C<0,$ $D<\delta_{2}$,
respectively, and the two resonances are always out of phase. The
typical spectra in zones $VII$ and $\uppercase\expandafter{\romannumeral10}$
are Fig. 2(c) and (d) with $C>0$ and $C<0$, respectively.

In the non-degenerate regime (upon the black dashed line), there are
six zones divided by the blue line, red line and the blue dash-dotted
line, and all blue dots  satisfy $C>0$. The blue line represents
$D=\gamma_{s},$ in which case the coefficient of $\chi_{E}$ vanishes
and the spectrum is the sum of two Lorentzian peaks {[}Fig. 3(c){]}.
The red line corresponds to $\Delta_{s}=\gamma_{s}$, indicating that
the splitting between the two resonances equals the linewidth. In
the zone $I$, we have $D<\gamma_{s},$ $\varDelta_{s}>\gamma_{s},$
the destructive interference between the ATS and EIT components leads
to enhanced transparency window, as shown in Fig. 3(b). When $D>\gamma_{s},$
$\varDelta_{s}>\gamma_{s},$ it is the zone $II$, the transparency
window of two Lorentz peaks is weakened due to the constructive interference
between ATS and EIT, as shown in Fig. 3(d). In the zone $III$, with
the typical spectrum shown in Fig. 3(a), the splitting between two
Lorentz peaks is smaller than the linewidth. The lineshape shows a
transparency window from EIT in Fig. 2(a) to the spectrum in Fig.
3(a, b), which illustrates the crossover from zones $V$ to $I$ via
zone $III$. The zone $IV$ means the incoherent transition rate $\gamma_{mr}$
becomes larger compared to the control light $\Omega_{c}$, then the
electromagnetically induced transparency with respect to the amplification
should be realized, and the zones $\uppercase\expandafter{\romannumeral8}$
and $\uppercase\expandafter{\romannumeral9}$ are opposite to the
zones $II$ and $IV$, respectively, with $C<0$.

\emph{Conclusion.-} The dispersion and absorption of probe light in
the three-level system is studied with the coherent laser driving
and incoherent control. In the degenerate regime, EIT in respect to
absorption or amplification is due to the interference of two resonances
with different linewidths. In the non-degenerate regime, the cross-over
from EIT to ATS is possible by increasing the driving power or controlling
the incoherent transition rates. By varying both coherent and incoherent
parameters of the three-level system, rich quantum phenomena of light-matter
interaction can be exemplified.

{\em Acknowledgments.} This work was funded by National Natural
Science Foundation of China (11074244 and 11274295), 973project (2011cba00200).
LJ acknowledges support from the Alfred P Sloan Foundation, the Packard
Foundation, the AFOSR-MURI, the ARO, and the DARPA Quiness program.

\end{document}